# A Planar Scanning Probe Microscope

S. Ernst[1], D.M. Irber[1], A.M. Waeber[1], G. Braunbeck[1] and F. Reinhard[1,*]

[1] Walter Schottky Institut and Physik Department, Technische Universität München, Am Coulombwall 4, 85748 Garching, Germany

*friedemann.reinhard@wsi.tum.de

**Scanning probe microscopy (SPM) is traditionally based on very sharp tips, where the small size of the apex is critical for resolution. This paradigm is about to shift, since a novel generation of planar probes (color centers in diamond[1], superconducting sensors[2] and single electron transistors[3]) promises to image small electric and magnetic fields with hitherto inaccessible sensitivity. To date, much effort has been put into fabricating these planar sensors on tip-like structures. This compromises performance and poses a considerable engineering challenge, which is mastered by only a few laboratories.**

**Here we present a radically simplified, tipless, approach – a technique for scanning an extended planar sensor parallel to a planar sample at a distance of few tens of nanometers. It is based on a combination of far-field optical techniques[4,5] to measure both tilt and distance between probe and sample with sub-mrad and sub-nm precision, respectively. Employing these measurements as a feedback signal, we demonstrate near-field optical imaging of plasmonic modes in silver nanowires by a single NV center.**

**Our scheme simultaneously improves the sensor quality and enlarges the range of available sensors beyond the limitations of existing tip-based schemes.**

The rise of nanotechnology has been largely fueled by scanning probe microscopy, most prominently atomic force and scanning tunneling microscopy, which have provided a tool to image topography, friction and electronic structure with a resolution down to the atomic level[6,7]. Today, the imaging of quantities other than forces and electric transport properties stands out as a frontier in this field of research. Scanning electron transistors (SETs) are being used to image weak electric stray fields[3,8] while weak magnetic fields are accessible to scanning Hall bars[9], nitrogen-vacancy (NV) centers in diamond[1,10–12] and scanning superconducting interference devices (SQUIDs)[13–15].

These two latter sensors have, more recently, even imaged surface temperature[16] and thermal conductivity[17].

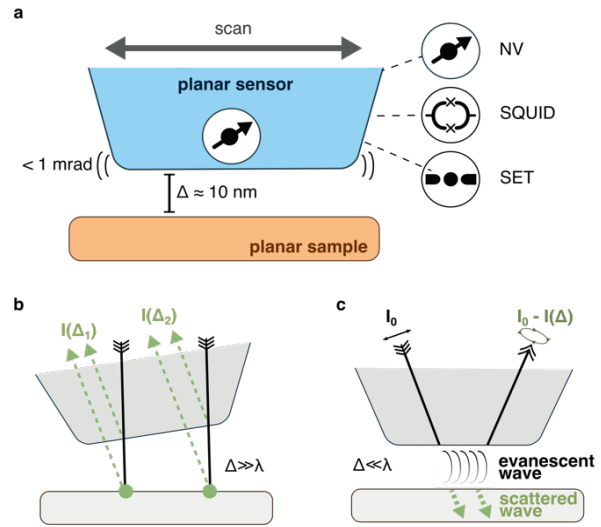

Fig 1 – Planar scanning probe microscope. **a**, An extended planar sensor (blue) can be brought into 10 nm-scale proximity with a planar sample (orange) if tilt is controlled to the sub-mrad level. NV: nitrogen-vacancy center, SQUID: superconducting sensor, SET: single electron transistor. **b**, Optical measurement of tilt by interference reflection microscopy. Reflections in a multilayer interfere, forming Newton's fringes. **c**, Optical measurement of the sensor-sample gap $\Delta$ by a combination of total internal reflection microscopy (TIRM) and Brewster angle microscopy (BAM). A sample scatters light from an evanescent field at the sensor surface, changing the intensity and/or polarization of the reflected beam. Effects of both **b** and **c** are imaged via a microscope looking from above.

Most of these novel techniques share a central challenge: the sensors are extended planar devices – circuits fabricated on a substrate or crystals – but scanning probe positioning requires a sharp tip to provide force feedback and to approach samples as close as possible. Existing approaches either resort to merely sliding the sensor over the sample[14], with obvious limitations on distance control and wear of the devices, or to fabricating the sensor onto a tip. Despite its technical complexity, this latter approach has been implemented for all major planar probes – for SETs and SQUIDs by sophisticated evaporation techniques[8,15] and for NV centers by anisotropic dry etching of diamond tips[1]. However, the considerable engineering effort involved restricts scanning probe experiments with these sensors to few highly specialized laboratories.

Nanofabrication also compromises sensor performance. Only the simplest SQUID geometry can be fabricated on tips – flux transformers, gradiometers and other elements that greatly enhance larger devices[14] cannot be used. The coherence time of NV centers typically degrades by orders of magnitude during nanofabrication, so that some of the most intriguing signals – such as nuclear magnetic resonance from nanoscale samples[18,19] – have only been detected with centers living under the planar surface of a bulk diamond so far.

Here we demonstrate a scanning probe microscope that enables imaging with extended planar sensors. It is based on the key idea (Fig. 1a) that such a sensor can actually be brought into nanometer-scale proximity with a sample if (i) the sample is also perfectly planar and (ii) sensor and sample can be aligned parallel to each other with sufficient precision. The first condition (i) is fulfilled for surprisingly many samples, including in particular cleaved crystals, thin films, and cryoslices of organic matter. The alignment condition (ii) could be ensured by nanopositioning if a means was found to measure tilt to provide a feedback signal for positioning. Similarly, a measurement of the sensor-sample distance could be fed back to stabilize the sensor-sample gap $\Delta$.

We find that two microscopy techniques are able to provide these measurements with sufficient precision. We measure tilt by interference reflection microscopy[4] (Fig. 1b). This technique effectively images Newton's rings forming between two partially reflective interfaces. Light rays reflected at different interfaces of a multilayer interfere, modulating the apparent brightness of the multilayer sample in a way that depends on the gap $\Delta$ ($I(\Delta)$ in Fig. 1b). Originally developed for studies of cellular membrane adhesion[4], the technique is widely used to measure the thickness of interface layers, such as the air cushion forming under a liquid droplet impacting on a surface[20–22], and has been combined with traditional atomic force microscopy to measure elasticity[23]. Applied to the Fresnel reflections from the sensor and sample surfaces it provides a tool to measure alignment, reaching a precision of few nanometers[24] over a surface of virtually arbitrary size.

While this technique could equally measure sensor-sample distance, we find that a small sensor-sample gap can be tracked with even higher sensitivity by a microscopy scheme reminiscent of total internal reflection microscopy (TIRM)[5,25] and Brewster angle microscopy (BAM, Fig. 1c)[26,27]. Here, we illuminate the sensor under an oblique angle larger than the critical angle for total internal reflection. An evanescent wave forms at the sensor surface, and a sample sufficiently close to the sensor absorbs or reflects part of its intensity. Absorption reduces the intensity of the reflected light while reflection alters its polarization (see below), generating an intensity contrast in a polarization microscope. Crucially, both effects are amplified by the short range of the evanescent field, which decays exponentially over a distance of ~100 nm. Similar techniques have been used widely to track Brownian motion of colloidal particles with nanometer precision[25] and to image lipid monolayers at the liquid-air interface[26,27].

We note that both techniques require the sensor to be translucent, and that the scanning probe microscope needs to be integrated into an optical microscope. This is a surmountable technical challenge for most sensors. In the case of diamond devices, which already need to fulfill both conditions for the readout of NV centers, it is even an experimental simplification. Here, scanning probe experiments can be performed in a slightly modified confocal microscope, obviating the need for a dedicated additional setup.

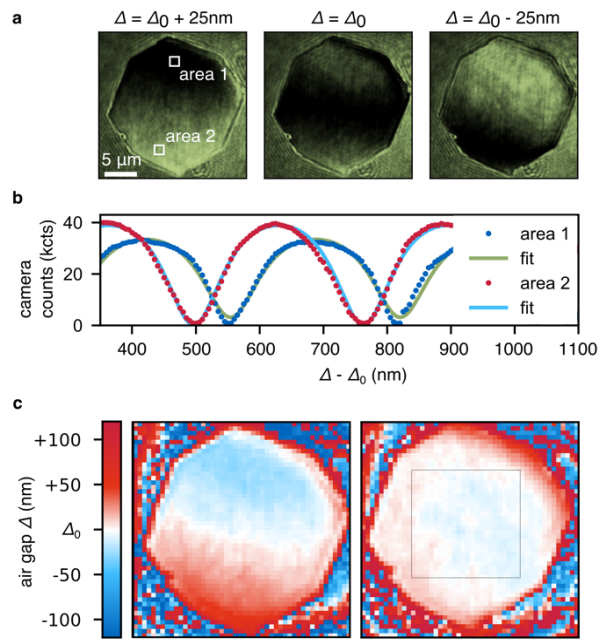

Fig 2 – Tilt alignment by interference reflection microscopy. **a**, A sample of polished silicon (see Methods), seen through a laterally millimeter-sized diamond sensor under collimated laser illumination along the optical axis. Interference fringes are visible as dark stripes. Upon approaching the sensor to the sample (images from left to right, decreasing $\Delta$), fringes move along the direction of steepest tilt. Tilt alignment is performed at an air gap of $\Delta_0 \approx 2$ μm. **b**, Determination of sensor-sample tilt. As the fringes move, the reflection intensity varies locally as an Airy distribution. A fit on the areas in **a** measures their relative displacement with a statistical error of 1.2 nm, equivalent to 0.1 mrad of tilt. **c**, Pixel-wise fits before alignment (left) and after alignment (right). The air gap varies by 22 nm peak-to-peak across the frame in (**c**, right) owing to systematic effects and residual tilt in the orthogonal direction.

We evaluate both schemes by approaching a 20 μm-sized sample of planar silicon with a laterally millimeter-sized bulk diamond sensor (Fig. 2, Methods). We use a comparatively small sample to lower the demands on alignment. This restriction could be circumvented by structuring the sensor on a 10 μm-scale instead, which would still be macroscopic compared to a nanoscale tip. The sample is approached to the sensor using a nano-positioner with five degrees of freedom (*xyz* translations and two rotation axes to correct for tilt). We perform imaging through the diamond sensor in a microscope under epi-illumination along the optical axis. When we approach the sample to a distance of a few microns or less, a cavity forms between the surfaces of the sensor and the sample. Maxima of cavity transmission, occurring at specific values of the sensor-sample air gap $\Delta$, become visible as dark stripes modulating the image. To quantitatively determine the sensor-sample tilt, we record reflection intensity while scanning the sample across a range of air gaps (Fig. 2b). Binned across a locally confined region, this intensity traces the Airy distribution of a Fabry-Perot interferometer. We fit this function over several minima, covering a range of a few microns, to obtain the displacement between different regions with an error of 1.2 nm, which corresponds to a measurement of tilt on the 0.1 mrad level. Repeating this analysis for every pixel of the image, we obtain a quantitative topographic map of $\Delta$ modulo $\lambda/2$ (Fig. 2c). We align the sensor and the sample by multiple iterations of measurement and subsequent manual tilt correction. This procedure reduces local variations of $\Delta$ to ~10 nm across the whole sample (Fig. 2c right), sufficiently small to approach a planar sensor as closely as tip-based scheme.

Once achieved, we find alignment to be sufficiently stable to perform the final approach and sample scan with control of the air gap $\Delta$ only. We measure $\Delta$ with sub-nanometer precision using a microscopy scheme combining total internal reflection illumination and Brewster angle microscopy (Fig. 3). We illuminate the sensor under an oblique angle larger than the critical angle of the diamond-air interface. Total internal reflection at the sensor surface rotates polarization, so that the illuminated sensor becomes visible as a bright surface in a polarization microscope (Fig. 3b). We maximize this effect by employing linearly polarized illumination, rotated by 45° to the plane of incidence to create a coherent superposition of s- and p-polarization. As we approach the sample into the evanescent wave region, it becomes visible as a dark area (Fig. 3b).

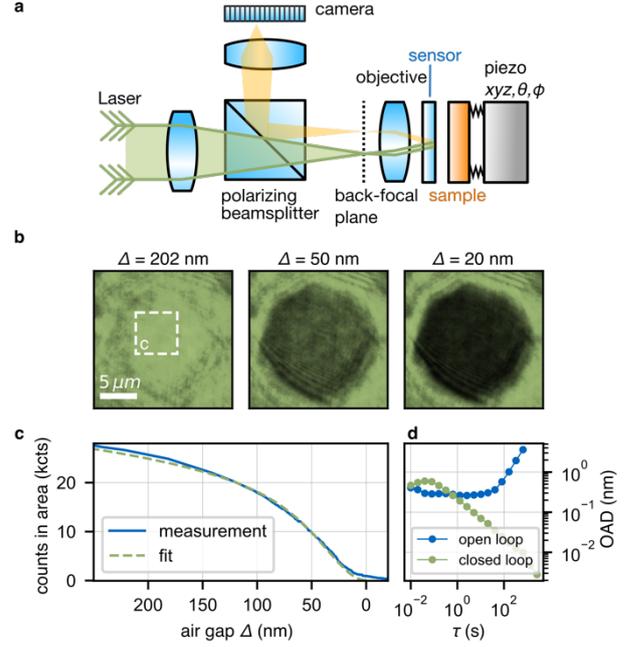

Fig 3 – Distance control. **a**, Experimental setup. An oblique illumination beam is internally reflected at the sensor surface. Reflection changes polarization, so that the reflected beam becomes visible in a polarization microscope. **b**, Microscope images of the same sensor and sample as Fig. 2. The sensor surface appears as a bright background. Once approached into the evanescent wave region, the sample appears as a dark spot. **c**, Fit of **b** to a model with two parameters (illumination angle θ = 25.49° ± 0.02°, predicted contact at $\Delta$ = 0.0 ± 0.4 nm). Departure from the fit for $\Delta \lesssim 20$ nm implies that sensor-sample contact occurs already at this point, owing largely to residual tilt. **d**, Overlapping Allan deviation (OAD) of the air gap. The gap is measured with sub-nanometer precision within 10 ms and can be stabilized by closed-loop control on a similar timescale.

We attribute this effect to the fact that reflection at the sample – typically a denser dielectric medium than air or a metallic surface – conserves polarization, so that light reflected at the sample becomes invisible in the polarization microscope. This interpretation is supported by theory, more specifically an exact analytical calculation of the reflection intensity by the transfer-matrix method (Fig. 3c). Loss of brightness upon approach is described excellently by this model using only two fitting parameters (illumination angle θ = 25.49° ± 0.02°, predicted contact at $\Delta = 0.0 \pm 0.4$ nm). This implies in particular that our method can measure absolute distance, since a fit obtained at large gaps accurately predicts the point of contact. We see a departure from the fit at very small distance ($\lesssim 20$ nm), suggesting that contact occurs already at this point. Piezo translation beyond this point (negative Δ in Fig. 3c) results in a less steep change in intensity, as sensor and sample are gradually squeezed together. This point of contact (at $\Delta \approx 20\,\text{nm}$) is consistent with our estimate of residual tilt (Fig. 2c) and an independent measurement of the minimum sensor-sample distance (Supplementary Information). Importantly, this value is comparable to the standoff distance achieved in tip-based schemes[12]. Brightness changes upon displacement are strong, with full contrast building up within only 150 nm. This is owing both to the high sensitivity of polarization-based schemes, reminiscent of ellipsometry and Nomarski microscopy, and to an additional amplification by the exponential decay of the evanescent wave. A measurement of $\Delta$ with sub-nanometer accuracy is obtained from a single camera image (Fig. 3d), which can be acquired within 10 ms or less. This is comparable to the typical pixel dwell-time in a scanning probe microscope, and hence sufficiently fast to employ the measurement as a feedback signal. We have stabilized the air gap in this manner by simple software-based feedforward control, achieving sub-nanometer stability over all timescales longer than 10 ms.

We finally turn to scanning probe experiments, using single NV centers in the diamond sensor as an optical near-field probe.

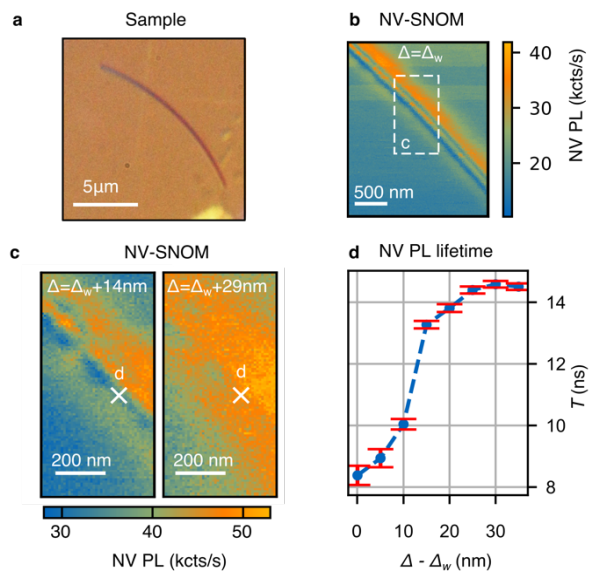

Fig 4 – Scanning Near-Field Optical Microscopy (SNOM) on Ag nanowires (diameter $\Delta_w \approx 60\,\text{nm}$) on a silicon substrate by a single NV center. **a**, Differential Interference Contrast (DIC) microscope image of the nanowire sample. **b**, Photo-Luminescence (PL) of a single NV center, scanned across the nanowire of **a** in contact mode. Quenching of luminescence becomes visible as a dark stripe of sub-diffraction width. **c**, Same as **b** for varying sensor-sample gap. Quenching disappears on the 10 nm scale. **d**, Fluorescence lifetime of a NV center hovering at varying distance above the wire. Data has been taken at the position highlighted in **c**.

All experiments are performed in the geometry of Fig. 1a. On a Si surface, we image Ag nanowires (Fig. 4a), which are known to support plasmonic modes that can quench the photoluminescence of NV centers in their 10 nm-scale vicinity[28]. These features are revealed clearly as we scan a NV center in a bulk diamond across the nanowire at the closest possible distance, maintaining a soft contact between the diamond and the sample (Fig. 4b). Features are highly sensitive to distance and disappear in scans at a height of only 30 nm above the wire surface (Fig. 4c). We measured the NV center fluorescence lifetime at varying height above the wire, finding a strong reduction in its proximity (Fig. 4d).

We expect our method to find widespread adoption, and its future impact to be larger than a mere technical simplification. First, the absence of nanofabrication will enable the use of high-quality sensors, such as NV centers with much longer coherence times. This will pave the way to routine imaging of weak signals such as nanoscale currents or nuclear spin ensembles. Second, it will enable a much wider range of sensors to be explored without the technical overhead of tip fabrication. We believe specifically that two-dimensional materials and their defects, as well as extended superconducting circuits, appear most promising but anticipate that many more avenues will develop with the new tool at hand.

# Methods

Experiments were performed in a homebuilt confocal microscope, upgraded by a 5D ($xyz\theta\phi$) slip-stick nano-positioner (SmarAct) for movement of the sample. Silicon samples were fabricated from a planar polished wafer (PLANO GmbH, nominal roughness 2-3 Å), isolating a protruding ~20 µm region of interest (Fig. 2a) by anisotropic etching with KOH. The diamond sensor was a 50 µm thick and millimeters wide scaife-polished diamond membrane, doped with single NV centers at a nominal depth of 10-20 nm beneath its surface (5 keV N$^+$ implant). Scanning probe imaging experiments (Fig. 4b,c) were performed in open loop, employing the distance control scheme as a mere monitor to verify that drift of $\Delta$ remained on a level of less than 5 nm. Fluorescence lifetimes were measured by excitation with a pulsed laser diode, time correlated detection, and a monoexponential fit of all photons arriving later than 10 ns after the end of the excitation pulse.

## Data availability

All relevant data is available from the authors upon request

## Acknowledgements

This work has been supported by the Deutsche Forschungsgemeinschaft via Emmy Noether grant RE3606/1-1, SPP1601 and the Nanosystems Initiative Munich (NIM). The authors wish to thank Michael Kaniber for helpful discussions and Christoph Hohmann for support with visualization.

## Author contributions

S.E. and F.R. conceived the measurement scheme. S.E. designed and built the experimental setup and fabricated the samples, with contributions from G.B. and A.M.W. Experiments were performed by S.E. and D.M.I, supervised by F.R.; S.E. and F.R. wrote the paper. All authors commented on the manuscript.

## Competing financial interests

The authors declare not to have competing financial interests.

## Material & Correspondence

Please send requests and correspondence to friedemann.reinhard@wsi.tum.de

# A Planar Scanning Probe Microscope

S. Ernst[1], D.M. Irber[1], A.M. Waeber[1], G. Braunbeck[1] and F. Reinhard[1,*]

[1] Walter Schottky Institut and Physik Department, Technische Universität München, Am Coulombwall 4, 85748 Garching, Germany

*friedemann.reinhard@wsi.tum.de

## Measurement of the minimum achievable sensor-sample distance

We calibrate the minimum achievable sensor-sample distance in an independent measurement using NV centers (Fig. S1). Specifically, we measure the standoff distance $\Delta_{NV}$ of an NV center beneath the sensor surface to the sample. This generally differs from the air gap $\Delta$, because of residual tilt, and since the centers are embedded slightly beneath the diamond surface (nominal depth 10-20 nm, 5 keV $N^+$ implant). To measure $\Delta_{NV}$ we employ standing waves, which form as the laser employed for NV excitation reflects from the sample surface, as an optical ruler (Fig. S1a). We determine $\Delta_{NV}$ by fitting the standing wave modulation at large $\Delta$, where additional modulation by Purcell enhancement is negligible, and extrapolate the fit to the intensity minimum closest to sensor-sample contact. Since this point can be equated with the node forming at the sample surface, the position of this minimum yields an estimate of $\Delta_{NV}$.
Performing this analysis on a series of NV centers yields estimates of $\Delta_{NV}$ between $-25$ nm and $+30$ nm, negative values arising from uncertainties of the fit and the identification of the contact point. This value is comparable to the standoff distance achieved in state-of-the-art experiments with tip-based schemes, typically around 50 nm[1].

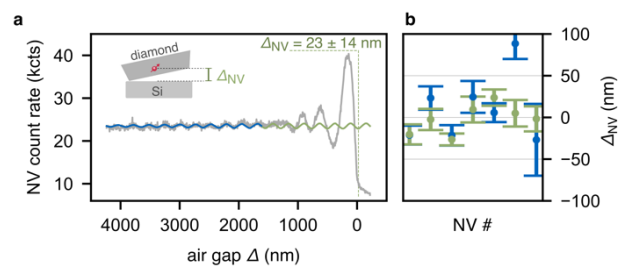

Fig. S1 – Measurement of the minimum achievable sensor-sample distance. **a**, Measurement of the standoff distance $\Delta_{NV}$ of an NV center embedded a few nanometers beneath the sensor surface. Fluorescence (grey curve) is modulated upon approach of a silicon surface – at short distances because of Purcell enhancement, at long distances due to standing waves in the excitation beam. We determine $\Delta_{NV}$ by fitting the standing waves (blue curve) and extrapolating the fit (green curve) to the node closest to contact, the position of which is $-\Delta_{NV}$. Contact is identified from a discontinuity in the fluorescence data and defined as $\Delta = 0$ nm. **b**, Measurement of standoff distance for several NV centers (blue: approach, green: retract). Negative values arise from the uncertainty in fitting the data and identifying contact. Centers can approach samples to within less than 50 nm, comparable to the distance achieved in tip-based schemes.